# On the clustering of rare codons and its effect on translation

Lalit Ponnala
Computational Biology Service Unit, Cornell University, Ithaca NY 14853
E-mail: lp257@cornell.edu

**Abstract**

The presence of clusters of rare codons is known to negatively impact the efficiency and accuracy of protein production. In this paper, we demonstrate a statistical method of identifying such clusters in the coding sequence of a gene. Using *E. coli* as our model organism, we show that genes having denser clusters tend to have lower protein yields.

*Keywords*: codon, translation, protein

**Introduction**

It is a common practice in biotechnology to substitute rare codons with more abundant ones, to enhance the yield of recombinant protein expression systems [1, 2]. Codons that are rarely used in the gene coding sequences have been shown to have low tRNA isoacceptor concentrations, and their excessive presence reduces the accuracy and efficiency of protein production [3].

Several studies have examined the effect of rare codons on heterologous gene expression, leading to many interesting findings. The presence of rare codons has been shown to modulate gene expresssion [3, 4] and attenuate viruses at the genome-scale [5]. More recently, rare codons have been found to affect the solubility of recombinant proteins [6] and give protein domains time to fold [7]. It is therefore of interest to know if such rare codons occur in clusters along a gene, and if a well-defined functional purpose can be attributed to their presence.

**Methods**

We use Kulldorff's spatial scan statistic to detect clusters of rare codons [8, 9]. The codons we mark as rare are those listed in a recent study of rare-codon function [7]:
(CUA, UCC, UCA, CCU, CCC, CCA, ACA, AGG, UUA, GUC)

Among the annotated coding sequences of *E. coli* genes (GenBank accession: NC_000913, http://www.ncbi.nlm.nih.gov/Genbank/), we assign binary values to each codon (1=rare, 0 otherwise) and use self-written MATLAB code (see http://sites.google.com/site/isbpaper/) to identify clusters on a gene-by-gene basis.

The relevant mathematics is as follows:
Let C be the number of rare codons in a gene of length N codons. Let c(Z) be the number of rare codons in a zone Z (a portion of the gene) containing n(Z) codons. The null hypothesis that no cluster exists in Z can be stated as $\frac{c(Z)}{n(Z)} = \frac{C-c(Z)}{N-n(Z)}$, i.e. the proportion of rare codons in zone Z is same as that outside Z. Let L(Z) be the likelihood under the alternative hypothesis that Z contains a cluster of rare codons, and let $L_0$ be the likelihood under the null hypothesis. In order to find the most likely cluster, the ratio $\lambda = \frac{L(Z)}{L_0}$ needs to be maximized over all zones Z. In our implementation, we choose zones centred on a rare codon and increment the zone-width linearly in steps of one codon in both directions.

Under the assumption that the distribution of rare codons follows a Bernoulli model, the likelihood ratio is shown to be [8]:

$$\lambda = \frac{\left(\frac{c(Z)}{n(Z)}\right)^{c(Z)} \left(1 - \frac{c(Z)}{n(Z)}\right)^{(n(Z)-c(Z))} \left(\frac{C-c(Z)}{N-n(Z)}\right)^{(C-c(Z))} \left(1 - \frac{C-c(Z)}{N-n(Z)}\right)^{((N-n(Z))-(C-c(Z)))}}{\left(\frac{C}{N}\right)^{C} \left(1 - \frac{C}{N}\right)^{(N-C)}}$$

when $\frac{c(Z)}{n(Z)} > \frac{C-c(Z)}{N-n(Z)}$ and 1 otherwise.

Significant clusters are found by creating Monte-Carlo replications of the rare-codon distributions, as described in [8].

This method is known to have high power, and performs well in practice, based on a comparison with other methods [10, 11].

**Results**

We looked for statistically significant rare-codon clusters (*P*-value<0.05) among genes of size 200-800 codons using Monte-Carlo simulation. All the 2716 genes we examined contained atleast one rare codon, and on an average, 8% of each gene was made up of rare codons.

We found roughly 13% of genes to have at least one significant cluster. Genes containing multiple clusters had an average gap of 143 codons between clusters. Genes of longer length (>300 aa) were found to be more likely to contain a cluster of rare codons. About one-third of all detected clusters occur in shorter genes (<300 aa). Among the detected clusters, about 23% were contained near translation starts, i.e. within the first 50 amino acids. We also found that the percentage of cluster-containing genes is roughly the same across various physiological roles.

The largest cluster (spanning 217 codons and containing 34 rare ones) was found in the gene *oppA*, which functions as an oligopeptide transporter subunit. The density of a cluster Z can be defined as the ratio of the number of rare codons in the cluster to the size of the cluster, i.e. d(Z)=c(Z)/n(Z). We found clusters of highest density (d(Z)=1) in 186 genes.

In order to check the validity of this method, we examined the mRNA-protein conversion efficiency (referred to as "yield") for genes that were found to contain significant clusters of rare codons. We used experimentally measured mRNA expression levels and protein abundances from Church Lab [12, 13].

We define yield as the ratio of protein abundance to mRNA expression level. We found a significant negative correlation between the yield of a gene and the density of clusters detected in it (correlation coefficient = -0.69, p-value = 0.038).

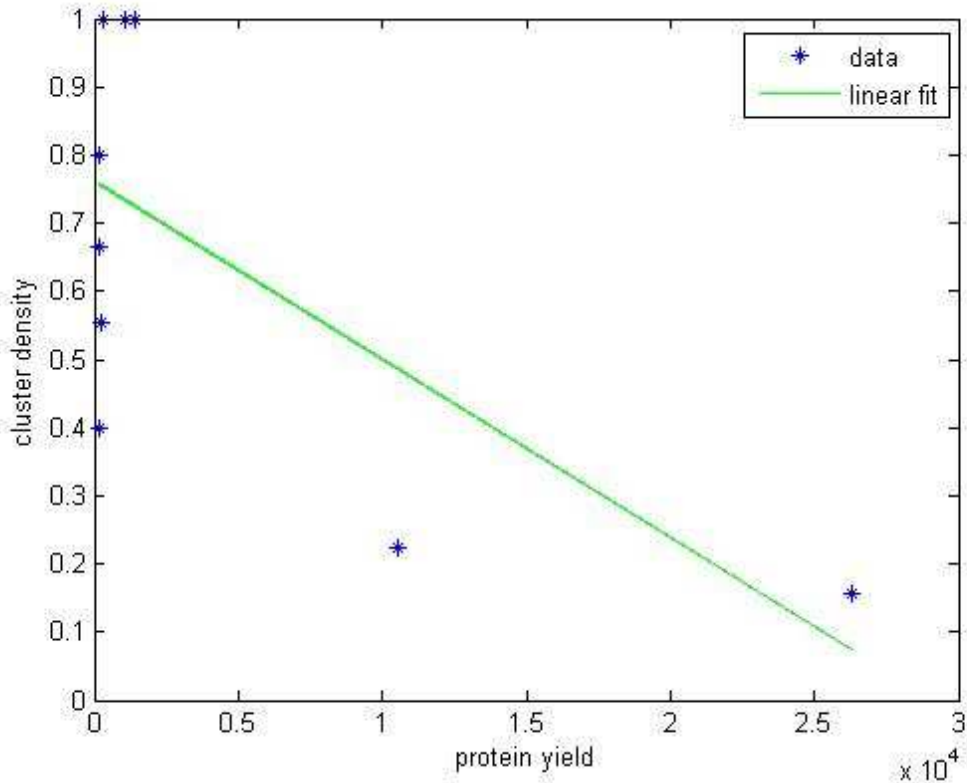

**Figure 1**: A plot of cluster density against protein yield showing negative correlation

**Discussion**

Several methods have been developed to identify rare codons in a gene sequence, many of which are publicly available via web-interfaces (see for example: http://nihserver.mbi.ucla.edu/RACC/ , http://molbiol.edu.ru/eng/scripts/01_11.html and http://www.doe-mbi.ucla.edu/~sumchan/caltor.html).  But relatively few methods help detect clusters of rare codons, and those that do [7, 14] use moving-averages to identify regions of the gene sequence where the codons have low tRNA concentration. Such methods seem to work in practice, but fail to take advantage of the wealth of statistical tools available for identifying clusters. They also tend to have a higher false-positive rate, owing to heuristic setting of the threshold for cumulative tRNA isoacceptor concentration.

The scan-statistic method developed by Kulldorff [8] seems well-suited for detecting the presence of clusters in spatially distributed data such as rare codons in a gene. Its only drawback is that it works in a "relative" way, i.e. the clusters it detects are significant compared to the rest of the gene being examined. In absolute terms, there may be many rough patches (small rare-codon clusters) in a gene, but many of them would not turn out to be significant via the scan-statistic approach. In a way, this works to our advantage because the genes in which we detect significant rare-codon clusters are more likely to have genuine translation problems. As we have shown, the clusters detected using this method seem to be biologically meaningful, since genes containing denser clusters of rare codons are found to have lower protein yield.